\documentclass[prl,twocolumn,superscriptaddress,showpacs]{revtex4}

\usepackage{graphicx}
\usepackage{ulem}
\usepackage{dcolumn}
\usepackage{bm}

\begin{document}

% single column 88 mm
% double column 170 mm

\title{Dynamic binding of driven interfaces in coupled ultrathin ferromagnetic layers}

\author{P.J.~Metaxas}
\email{metaxas@physics.uwa.edu.au}
\altaffiliation{Present address: Unit\'e Mixte de Physique CNRS/Thales, 1 Ave.~A.~Fresnel, 91767 Palaiseau Cedex, France.}
\affiliation{School of Physics, M013, University of Western Australia, 35 Stirling Hwy, Crawley WA 6009, Australia.}
\affiliation{Laboratoire de Physique des Solides, Universit\'e Paris-Sud, CNRS, UMR 8502, 91405 Orsay Cedex, France.}

\author{R.L.~Stamps}
\affiliation{School of Physics, M013, University of Western Australia, 35 Stirling Hwy, Crawley WA 6009, Australia.}

\author{J.-P.~Jamet}
\author{J.~Ferr\'{e}}
\affiliation{Laboratoire de Physique des Solides, Universit\'e Paris-Sud, CNRS, UMR 8502, 91405 Orsay Cedex, France.}

\author{V.~Baltz}
\author{B.~Rodmacq}
\affiliation{SPINTEC, UMR-8191, CEA-INAC/CNRS/UJF-Grenoble 1/Grenoble-INP, 17 rue des Martyrs, 38054 Grenoble cedex 9, France.}

\author{P.~Politi}
\affiliation{Istituto dei Sistemi Complessi, Consiglio Nazionale delle Ricerche, Via Madonna del Piano 10, 50019 Sesto Fiorentino, Italy.}
\date{\today}
\pacs{75.70.Cn, 75.60.Ch, 75.78.Fg}
%75.60.Ch	Domain walls and domain structure (for magnetic bubbles and vortices, see 75.70.Kw)
%75.78.Fg	Dynamics of domain structures
%Magnetic multilayers, 75.70.Cn

\begin{abstract}
We  demonstrate experimentally dynamic interface binding in a system consisting of two coupled ferromagnetic layers. While domain walls in each layer have different velocity-field responses, for two broad ranges of the driving field, $H$, walls in the two layers are bound and move at a common velocity. The bound states have their own velocity-field response and arise when the isolated wall velocities in each layer are close, a condition which always occurs as $H\rightarrow 0$. Several features of the bound states are reproduced using a one dimensional model, illustrating their general nature.
\end{abstract}

\maketitle

Moving elastic interfaces are a common feature of many non-equilibrium phenomena. These include  crystal growth \cite{Drossel2000}, wetting \cite{Moulinet2002,Balankin2006}, combustion \cite{Zhang1992,Myllys2001},  vortex motion in superconductors \cite{Blatter1994},  mechanical fracturing \cite{Ponson2006} and switching in ferromagnetic and ferroelectric materials \cite{Lemerle1998,Metaxas2007,KrusinElbaum2001,Kim2009,Paruch2006}. However, the majority of work examining the  universal dynamics of these interfaces has  been limited to studies of single interfaces, and it is only quite recently that the problem of coupled interfaces  has begun to be addressed \cite{Barabasi1992,Majumdar2005,Juntunen2007}.

Up until now, experimental studies have been restricted to the particular problem of two interfaces moving through a single medium generally while separated by a finite lateral distance \cite{Balankin2006,Bauer2005}. In this Letter  we consider interface dynamics in a novel type  of experimental system consisting instead of two coupled, but physically separate, media. The interfaces, magnetic domain walls, move through two ferromagnetically coupled ultrathin ferromagnetic layers under the action of an applied driving field, $H$.  The central result of this work is clear evidence of a dynamic binding of the domain walls in the two media for certain finite ranges of $H$.  Due to the two layers having different  disorder strengths and thicknesses \cite{Metaxas2007},   in the absence of coupling  the domain walls in each  layer have different velocity-field responses. Experimentally however, we find that at two certain values of the driving field, $H=0$ and $H=H^*>0$, the velocities in each layer are the same. We show that due to the interlayer coupling, a dynamic binding of the walls appears over finite  ranges of the applied field near the crossing points, $H=0$ and $H^*$. In these field ranges, walls in the separate layers move together at a common velocity. Notably, the bound states are characterized by their own velocity-field response, many features of which can be reproduced by a one dimensional  model which takes the strength of the interlayer coupling into account.

\begin{figure}
		\includegraphics[width=7cm]{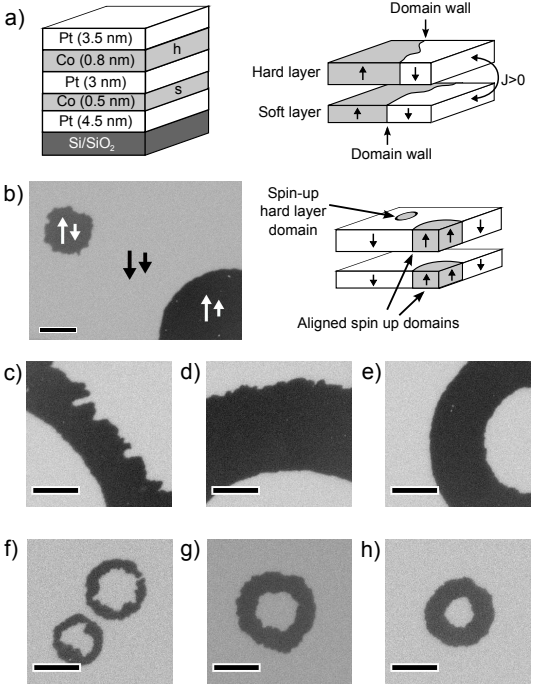}
		\caption{(a) `Hard' (h) and `soft' (s) Co layers, sandwiched between Pt layers, interacting via a ferromagnetic interlayer coupling, $J$. Domain walls separate `spin up' and `spin down' domains. 
(b) PMOKE image of magnetic domains in the hard and soft layers showing the relative orientations of the magnetization in each domain: large (small) arrows for the hard (soft) layer. `Spin up' domains %are surrounded by `spin down' domains and 
may either be isolated  in the hard layer (mid-level gray) or exist in both layers (black). Subtracted  PMOKE images for aligned (c-e) and hard layer (f-h) walls (dark regions correspond to areas swept out by the domain walls). Field pulse amplitudes and durations are: (c) 112 Oe 367 s (d) 638 Oe 3$\times$700 ns (e) 1120 Oe 250 ns (f) 155 Oe 45 ms (g) 670 Oe 500 ns (h) +1040 Oe 2$\times$80 ns. Black scale bars are 10 $\mu$m long in images (b-h).
}
\end{figure}

Our system, shown in Fig.~1(a), consists of two ultrathin weakly disordered ferromagnetic Co layers with perpendicular anisotropy which interact via a net ferromagnetic (FM) interlayer coupling \cite{Moritz2004} of energy $J$. Sandwiched between Pt, such ultrathin Co layers are well established as model systems for studying the dynamics of one dimensional (1D) interfaces  moving through a weakly disordered two dimensional (2D)  medium (walls are narrow $\sim$10 nm) \cite{Lemerle1998,Metaxas2007,KrusinElbaum2001,Kim2009}. The multilayer stack, having structure Pt(4.5 nm)/Co(0.5 nm)/Pt(3 nm)/Co(0.8 nm)/Pt(3.5 nm) was sputtered at room temperature onto an in-situ etched Si/SiO$_2$ substrate. The two Co layers switch together during hysteresis, consistent with an FM coupling \cite{note1}.
The magnetically `hard' thick 0.8 nm Co layer has a stronger depinning field as compared to the `soft' 0.5 nm Co layer which, as will be seen below, results in slower field induced domain wall propagation at low field.

Domain wall velocities, $v(H)$, with $H$  applied perpendicular to the film plane, were determined quasi-statically from high resolution (0.4 $\mu$m) far-field polar magneto-optical Kerr effect (PMOKE) microscopy images.  Both magnetic layers were first saturated in a strong negative field ($|H|=1$ kOe to $4$ kOe). Short positive field pulses ($\sim$1 kOe over $\sim$100 ns) could then be used to nucleate isolated `spin up' domains in the hard layer and aligned `spin up' domains in both layers [Fig.~1(b)] \cite{note3}.  These latter aligned domains  allow us to study bound domain wall dynamics.  Domain configurations were imaged both before and after the application of a second field pulse applied to drive wall motion. These two images were  subtracted from each other [eg.~Figs.~1(c-h)] and the average  wall displacement and  velocity were determined. Further details regarding this method can be found in Ref.~\cite{Metaxas2007}.

The characteristic field dependence  of the velocities of isolated hard and soft domain walls, $v_h$ and $v_s$, in the absence of any interlayer coupling are plotted in Fig.~2(a).
In both layers, wall motion at low field is consistent with a thermally activated creep regime \cite{Lemerle1998,Chauve2000}: 
\begin{equation}
%v=v_0 \exp \left[-\frac{U_C}{k_B T} \left( \frac{H_{dep}}{H} \right)^{1/4} \right]
v(H)=v_0 \exp \left[-\frac{U_C}{k_B T}\left(\frac{H_{dep}}{H}\right)^{1/4} \right]
\label{e:creep}	
\end{equation}
as demonstrated  in the $\ln v_{h,s}(H^{-1/4})$ plot  in Fig.~2(b).
In Eq.~(1), $U_C/k_BT$ is related to the disorder-induced pinning energy barriers, $v_0$ is a numerical prefactor and $1/4$ is the universal dynamic exponent for a 1D interface moving in a 2D weakly disordered medium. The slope of  $\ln v\left(H^{-1/4}\right)$ is equal to $-H_{dep}^{1/4}U_C/k_BT$. It increases with the disorder strength and is higher for the thicker hard layer \cite{Metaxas2007}. $v_h$ was measured directly  in the hard layer of this film. However, since we could not nucleate isolated soft layer domains, $v_s$ was measured in the soft 0.5 nm Co layer of a sample with a 4 nm spacer (resulting in a weak antiferromagnetic interlayer coupling \cite{Metaxas2008}) but an otherwise  identical structure to that of the film studied here.

 \begin{figure}
		\includegraphics[width=8cm,clip=true]{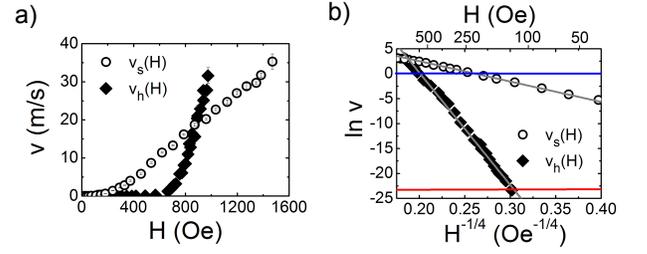}
		\caption{(Color online) Isolated domain wall velocities in the hard ($v_h$) and soft ($v_s$) layers in the absence of 	coupling plotted as (a) $v(H)$ and (b) $\ln v(H^{-1/4})$.}
\end{figure}

We  now discuss how ferromagnetic interlayer coupling may theoretically give rise to bound states. Existence of bound states relies on two important features of our system. The first feature is that the $v_h(H)$ and $v_s(H)$ curves [Fig.~2(a)] cross  at two field values $H=H^*\approx 870$ Oe and at $H=0$. Note that this second crossing point is universal  since $v\rightarrow 0$ as $H\rightarrow 0$. The second feature, as illustrated in Fig.~3(a), is that the ferromagnetic interlayer coupling induces an attractive interaction between walls in each layer. This is mediated by effective coupling fields, $H_J^i=J/(t_iM^i_S)$ where $M_S^i$ and $t_i$ are the saturation magnetization and  thickness of layer $i=h,s$   \cite{Grolier1993-2}.  These fields  act  to align the magnetization of the domains in each layer and hence the domain walls (eg.~\cite{Robinson2006}).

\begin{figure}
		\includegraphics[width=7cm]{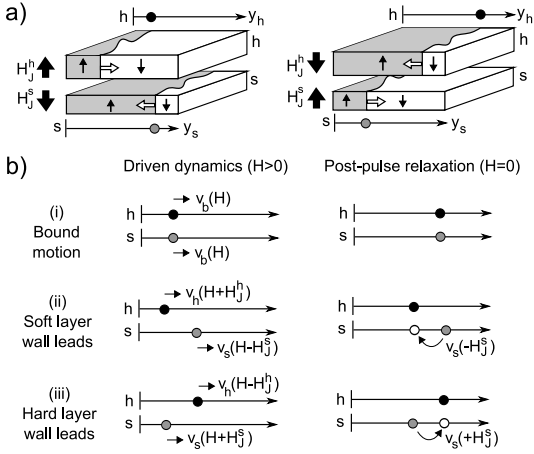}
		\caption{ (a) The sign of the coupling fields, $H_J^{s,h}$, is such that walls in the two layers are driven towards one another (white arrows).  Positive applied fields will tend to drive each wall to the right and in the 1D model  the average positions of the walls (filled circles) are considered. (b) The total field acting on the walls during driven motion is $H\pm H_J^{h,s}$ ($H_J^{h,s}= 0$ for aligned walls). For certain field ranges neither wall can move faster than the other, resulting in a dynamic bound state with $v_h=v_s=v_b$ (i). Outside these field ranges, unbound motion will occur with $v_s>v_h$ (ii) or $v_h>v_s$ (iii). In these latter cases, when the applied field is removed, the soft layer wall will relax to the position of the hard layer wall under the influence of $\pm H_J^s$.
}
\end{figure}

Ferromagnetic (attractive) coupling therefore competes with the tendency of the domain walls to move separately. If $v_h\approx v_s$, the former prevails over the latter and a bound state is favored. On these very general grounds, from Fig.~2(a) we expect three regimes which are shown schematically in the left hand column of Fig.~3(b).  (i) For $H\approx 0$ and $H\approx H^*$, walls are bound and propagate together at a velocity $v_b(H)$. Far from these field values, walls move separately, either (ii) with $v_s>v_h$ or (iii) with $v_h>v_s$. Following Fig.~3(a), laterally separated walls will move under the action of a total field  \cite{Fukumoto2005,Metaxas2008}, $H\pm H_J^{s,h}$,   according to their relative
positions:
$H+H_J^{s,h}$ (increased $v$) for the trailing wall and $H-H_J^{s,h}$ (decreased $v$) for the leading wall.

Domain imaging was always carried out after removal of the driving field. Upon the removal of $H$, walls which are  separated during their field driven motion will relax  towards each other under the  action of an effective field equal to $\pm H_J^{s,h}$  as shown in the right column of Fig.~3(b). From the data in Fig.~2(b) (see the blue and red lines corresponding respectively to $\ln v_s(H_J^s)$ and $\ln v_h(H_J^h)$) we find that 
$v_s(H_J^s)\approx$ 1 m/s $\approx 10^{10}v_h(H_J^h)$, indicating that effectively only the soft layer wall will move significantly, relaxing rapidly  to the hard layer wall position in a time much shorter than the  $\sim$10 s needed for image acquisition. This results in an apparent displacement of the aligned walls corresponding to that of the hard layer wall.

In light of the above considerations we  now examine experimental results for the coupled dynamics in the low field regime. In Fig.~4(a) we plot $\ln v_a$ as a function of $H^{-1/4}$
 where $v_a$ is the  velocity of aligned walls as determined from their experimentally observed displacements.  The crucial feature  is a distinct change in slope at $H\approx 250$ Oe. 
Above this field, the measured $v_a$ dynamics correspond to those of hard layer walls driven under the action of a positive driving field and a positive $H_J^h$ (the $v_h(H+H_J^h)$ data). This is what is expected for unbound walls with  $v_s>v_h$, the situation shown in Fig.~3(b,ii).  Below this field, unique dynamics are observed with $v_a(H)\ne v_h(H\pm H_J)$. Wall motion here is shown below to be consistent with bound dynamics. Additionally, since  $\ln v_a\propto H^{-1/4}$, this regime is consistent  with a bound creep regime. Interestingly, the energy barriers for the bound walls appear to be defined by disorder in the hard layer (compare the slopes of the $\ln v_h\left( H^{-1/4}\right)$ and $\ln v_a\left( H^{-1/4}\right)$ data below 250 Oe) with the increased velocity for the bound walls consistent with a larger $v_0$ [see Eq.~(1)].

  \begin{figure}
		\includegraphics[width=7cm]{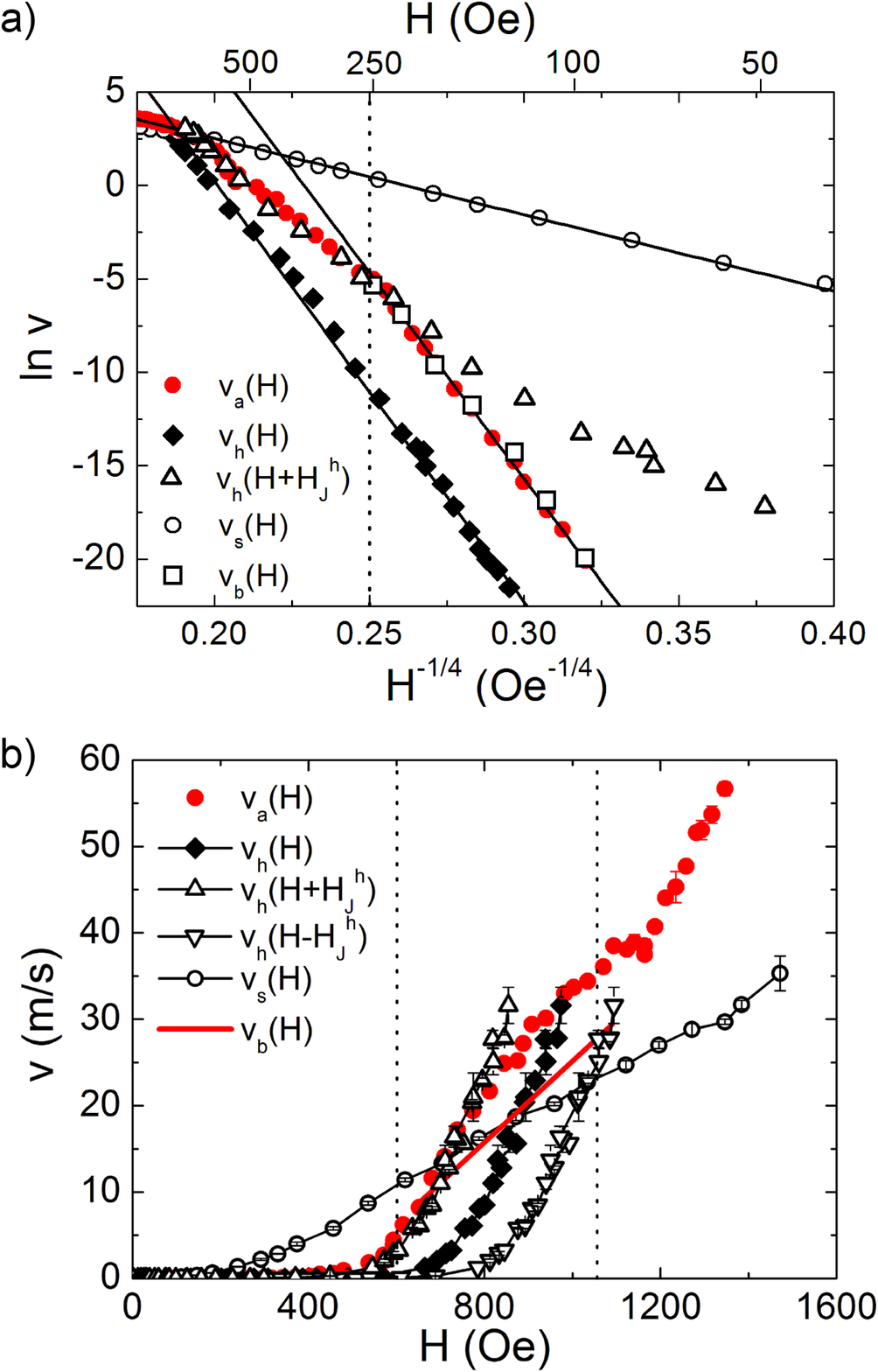}
		\caption{(Color online) $v_a(H)$ corresponds to dynamics determined experimentally from walls aligned at remanence. $v_h(H)$ and $v_s(H)$ are isolated hard and soft layer velocities in the absence of coupling [Fig.~2]. Hard layer dynamics in the presence of a positive coupling field ($v_h(H+H_J^h)$) and a negative coupling field ($v_h(H-H_J^s)$) are also shown.  Low field creep dynamics are shown in (a).  Fits to the linear low field $v_{h,s,a}(H)$ data are plotted as solid black lines.  $v_{b}(H)$ are bound wall velocities calculated using Eq.~(3) with experimentally determined coupling fields. Full velocity field curves are shown in (b). The calculated $v_{b}(H)$ for the high field bound state, also determined using Eq.~(3) is  shown as a red line. The vertical dotted lines correspond to the predicted lower and upper limits for the high field bound state.}
\end{figure}

Unbound motion above $H\approx 250$ Oe can be sustained only if $v_s(H-H^s_J)>v_h(H+H^h_J)$ [Fig.~2(b,ii)]. This occurs for $H>H_{c1}$ where
\begin{equation}
	v_s(H_{c1}- H^s_J)=v_h(H_{c1} + H^h_J).
	\label{e:cond}
\end{equation}
After fitting the low field data in Fig.~2 with Eq.~(1), we can use Eq.~(2)  to estimate  $H_{c1}=254$ Oe where we have used  experimentally determined values for the coupling fields: $H_J^h=120$ Oe and $H_J^s=220$ Oe \cite{note2}. This value for $H_{c1}$ agrees very well with the field at which a slope change is observed in the $\ln v_a(H^{-1/4})$ data in Fig.~4(a) (see the vertical dotted line at $H=254$ Oe). 
For $H<254$ Oe, the coupling fields  prevent either wall leading the other, thereby binding them [Fig.~3(b,i)]. 

The bound wall velocity can be predicted by modeling the coupling fields with
$h_J^{h,s}(d)= H_J^{h,s} \tanh(d/\Delta)$ where 
$d=y_s-y_h$ is the distance separating walls [Fig.~3(a)]. The out of plane coupling field is expected to be proportional to the out of plane component of the magnetization, $m_z$, and this model describes $m_z$ for a Bloch domain wall of width $\Delta$ (eg.~\cite{bubblematerials}) where we use the same $\Delta$ for both layers. In essence, $\Delta$ simply 
 determines the
length over which the coupling `force' changes sign.
In place of Eq.~(2), the condition for bound motion is then
\begin{equation}
	v_s(H- h^s_J(d))=v_h(H+ h^h_J(d)) \, .
\end{equation} 
We can numerically solve Eq.~(3) for $d$ using the $v_{s,h}(H)$ data in Fig.~2 together with $H_J^{s,h}$. If $d$ can be determined for a given $H$, $v_h=v_s=v_b$ and $v_b$ can be obtained  simply using Eq.~(3). Results are plotted in Fig.~4(a). Experiment agrees well with  the predicted $v_b$ for fields below 254 Oe where bound motion is expected. Note that our result does not depend on $\Delta$ (therefore giving a model with no free parameters) and that for a given field, no bound state exists if Eq.~(3) has no solution for $d$ (therefore providing an alternative method to find $H_{c1}$ which is consistent with that discussed above).

In Fig.~4(b), a second bound state, also with a unique velocity-field response, is identified around $H^*\approx 870$ Oe. Here, a change in slope of the $v_a(H)$ data is again observed due to a deviation away from the $v_h(H\pm H_J^h)$ unbound behaviour.
Away from this second bound regime, the velocity-field response of the aligned walls  compares well to that of unbound walls with the  hard layer wall either trailing ($H<840$ Oe, $v_a\approx v_h(H+H_J^h)$, Fig.~3(b,ii)) 
or leading ($H>1190$ Oe, $v_a\approx v_h(H-H_J^h)$, Fig.~3(b,iii)). We again use  Eq.~(3) to predict  this bound state's velocity  $v_{b}(H)$ and its limit fields, $H_{c2}\approx 600$ Oe and $H_{c3}\approx 1050$ Oe, all shown in Fig.~4(b). 
While our model appears to capture the essential physics of the stability and dynamics of the bound states, at high field it predicts a bound velocity which lies between the measured $v_h(H)$ and $v_s(H)$ whereas the observed bound velocity is  higher. 
We note that our simple calculation of the velocities does not allow for any effects that may appear in a full 2D treatment of the problem (eg.~elasticity) nor do we consider  dipolar fields generated at the domain walls \cite{Baruth2006}.

 In conclusion, we report the discovery of a dynamic binding of driven interfaces  with general characteristics that can be understood to be a consequence of attractive interactions and velocities
that  match at some  driving force. Since $v\rightarrow 0$ always for a vanishing force, the crossing point at $H=0$ is universal and there may
be analogies with  other systems, for example coupled vortex motion in superconductors \cite{White1991,Ustinov1996} and  bound solitons \cite{How1989}.

\begin{acknowledgments}
P.J.M. acknowledges the support of the Australian Government, a Marie Curie Early Stage Training Fellowship
(MEST-CT-2004-514307) and a UWA Whitfield Fellowship. P.J.M., R.L.S. and J.F. were supported by the French-Australian Science and
Technology (FAST) program and the `Triangle de la Physique' through the PRODYMAG research project. P.J.M., R.L.S. and P.P. acknowledge support from the Australian Research Council and the Italian Ministry of Research (PRIN 2007JHLPEZ). Thanks to A.~Mougin for a critical reading of the manuscript.
\end{acknowledgments}


\begin{thebibliography}{30}
\bibitem[]{note3}
A PMOKE signal is obtained from both layers due to the low thickness of the stack with respect to the penetration depth of the light. Higher contrast is obtained from the aligned domains as the signals from the hard and soft layer domains add constructively in these positions.
\end{thebibliography}
\end{document}